\documentclass[a4paper,twocolumn,11pt]{article}

\usepackage[left=2cm,right=2cm]{geometry}

\usepackage{amsfonts,amsmath}
\usepackage{verbatim}
\usepackage{listings}

\usepackage{blkarray}
\makeatletter
\let\BA@quicktrue\BA@quickfalse
\makeatother

\usepackage{dcolumn}
\usepackage{rotating}
\usepackage{graphicx}

\usepackage[hyperref=auto]{biblatex}
\usepackage{dmbiblatex}
\bibliography{lctes25-henry}

\usepackage[inline]{enumitem}

\usepackage{tikz}
\usetikzlibrary{arrows,shadows,shapes,backgrounds,fit,chains}
\usepackage{color}
\usepackage{hyperref}
\usepackage{cleveref}

\lstdefinelanguage{smtlib2}
{morekeywords={assert,check-sat,get-model,declare-fun,and,or,ite,not,true,false,Int,Bool,Real},
sensitive=false,
morecomment=[l]{;},
morestring=[b]",
}

\lstdefinelanguage{llvm}
{morekeywords={},
sensitive=false,
morecomment=[l]{;},
morestring=[b]",
}
\lstnewenvironment{llvm}
{\lstset{language=llvm,
		basicstyle=\fontfamily{ppl}\small,
		commentstyle=\textit,
		keywordstyle=\textbf,
		showstringspaces=false}}
{}

\newcommand{\mytitle}{How to Compute Worst-Case Execution Time by Optimization Modulo Theory and a Clever Encoding of Program Semantics}
\title{\mytitle{\renewcommand*{\thefootnote}{\fnsymbol{footnote}}\footnotemark}}

\author{Julien Henry \and Mihail Asavoae \and David Monniaux \and Claire Ma\"iza}

\hypersetup{pdftitle={\mytitle},pdfauthor={Julien Henry, Mihail Asavoae, David Monniaux, Claire Ma\"iza}}

\newcommand{\ite}{\mathit{ite}}
\newcommand\true{\mathit{true}}

\newcommand{\NN}{\mathbb{N}}

\newcommand{\RegTM}{\texttrademark}

\begin{document}
\maketitle
 
{\renewcommand*{\thefootnote}{\fnsymbol{footnote}}
\footnotetext[1]{The research leading to these results has received funding from the French \emph{Agence nationale de la recherche}, grant \href{http://wsept.inria.fr/}{W-SEPT} (ANR-12-INSE-0001), and from from the European Research Council under the European Union's Seventh Framework Programme (FP7/2007--2013) / ERC grant agreement 306595 ``\href{http://stator.imag.fr/}{STATOR}''.

This article was also published in the proceedings of the 2014 ACM SIGPLAN Conference on Languages, Compilers and Tools for Embedded Systems (LCTES).}}

\begin{abstract}
In systems with hard real-time constraints, it is necessary to compute upper bounds on the worst-case execution time (WCET) of programs; the closer the bound to the real WCET, the better.
This is especially the case of synchronous reactive control loops with a fixed clock; the WCET of the loop body must not exceed the clock period.

We compute the WCET (or at least a close upper bound thereof) as the solution of an \emph{optimization modulo theory} problem that takes into account the semantics of the program, in contrast to other methods that compute the longest path whether or not it is feasible according to these semantics.
Optimization modulo theory extends satisfiability modulo theory (SMT) to maximization problems.

Immediate encodings of WCET problems into SMT yield formulas intractable for all current production-grade solvers --- this is inherent to the DPLL(T) approach to SMT implemented in these solvers.
By conjoining some appropriate ``cuts'' to these formulas, we considerably reduce the computation time of the SMT-solver.

We experimented our approach on a variety of control programs, using the OTAWA analyzer both as baseline and as underlying microarchitectural analysis for our analysis, and show notable improvement on the WCET bound on a variety of benchmarks and control programs.

\end{abstract}

\section{Introduction}\label{sec:intro}
In embedded systems, it is often necessary to ascertain that the worst-case execution time (WCET) of a program is less than a certain threshold.
This is in particular the case for synchronous reactive control loops (infinite loops that acquire sensor values, compute appropriate actions and update, write them to actuators, and wait for the next clock tick) \parencite{Caspi_et_al_Synchronous_2008}: the WCET of the loop body (``step'') must be less than the period of the clock.

Computing the WCET of a program on a modern architecture requires a combination of low-level, microarchitectural reasoning (regarding pipeline and cache states, busses, cycle-accurate timing) and higher-level reasoning (program control flow, loop counts, variable pointers).
A common approach is to apply a form of abstract interpretation to the microarchitecture, deduce worst-case timings for elementary blocks, and reassemble these into the global WCET according to the control flow and maximal iteration counts using integer linear programming (ILP) \parencite{Theiling_et_al_2000,wilhelm-et-TECS08}.

One pitfall of this approach is that the reassembly may take into account paths that cannot actually occur in the real program, possibly overestimating the WCET.
This is because this reassembly is mostly driven by the control-flow structure of the program, and (in most approaches) ignores semantic conditions.
For instance, a control program may (clock-)enable certain parts of the program according to modular arithmetic with respect to time:
\begin{lstlisting}
if (clock % 4==0) { /* A */ }
/* unrelated code */
if (clock % 12==1) { /* B */ }
\end{lstlisting}
These arithmetic constraints entail that certain combinations of parts cannot be active simultaneously (sections $A$ and $B$ are mutually incompatible).
If such constraints are not taken into account (as in most approaches), the WCET will be grossly over-estimated.

The purpose of this article is to take such \emph{semantic constraints} into account, in a fully automated and very precise fashion.
Specifically, we consider the case where the program for which WCET is to be determined contains only loops for which small static bounds can be determined (but our approach can also be applied to general programs through summarization, see \autoref{sec:extensions}).
This is very commonly the case for synchronous control programs, such as those found in aircraft fly-by-wire controls \parencite{Souyris_et_al_FM2009}.
Programs of this form are typically compiled into C from high-level data-flow synchronous programming languages such as \textsc{Simulink}%
\footnote{\textsc{Simulink}{\RegTM} is a block diagram environment for multidomain simulation and model-based design from The Mathworks.},
\textsc{Lustre} or
\textsc{Scade}%
\footnote{\textsc{Scade}{\RegTM} is a model-based development environment dedicated to critical embedded software, from Esterel Technologies, derived from the academic language \textsc{Lustre}.}
\parencite{Caspi_et_al_Synchronous_2008}.

We compute the WCET of such programs by expressing it as the solution of an \emph{optimization modulo theory} problem. Optimization modulo theory is an extension of \emph{satisfability modulo theory} (SMT) where the returned solution is not just any solution, but one maximizing some objective; in our case, solutions define execution traces of the program, and the objective is their execution time.

Expressing execution traces of programs as solutions to an SMT problem is a classical approach in \emph{bounded model checking}; typically, the SMT problem includes a constraint stating that the execution trace reaches some failure point, and an ``unsatisfiable'' answer means that this failure point is unreachable.
In the case of optimization, the SMT solver has to disprove the existence of solutions greater than the maximum to be returned --- in our case, to disprove the existence of traces of execution time greater than the WCET.
Unfortunately, all currently available SMT solvers take unacceptably long time to conclude on naive encodings of WCET problems.
This is because all these solvers implement variants of the DPLL(T) approach \parencite{Kroening_Strichman}, which has exponential behavior on so-called ``diamond formulas'', which appear in naive encodings of WCET on sequences of if-then-elses.

Computing or proving the WCET by direct, naive encoding into SMT therefore leads to intractable problems, which is probably the reason why, to our best knowledge, it has not been proposed in the literature.
We however show how an alternate encoding, including ``cuts'', makes such computations tractable.
Our contributions are:

\begin{enumerate}
\item The computation of worst-case execution time (WCET), or an over-approximation thereof, by optimization modulo theory. The same idea may also be applicable to other similar problems (e.g. number of calls to a memory allocator).
Our approach exhibits a worst-case path, which may be useful for targeting optimizations so as to lower WCET~\parencite{DBLP:journals/rts/ZhaoKWHM06}.
\item The introduction of ``cuts'' into the encoding so as to make SMT-solving tractable, without any change in the code of the SMT solver. The same idea may extend to other problems with an additive or modular structure.
\end{enumerate}

In \autoref{sec:wcet}, we recall the usual approach for the computation of an upper bound on WCET. In \autoref{sec:bmc}, we recall the general framework of bounded model checking using SMT-solving. In \autoref{sec:cuts}, we explain how we improve upon the ``normal'' SMT encoding of programs so as to make WCET problems tractable, and in \autoref{sec:intractable} we explain (both theoretically and practically) why the normal encoding results in intractable problems. In \autoref{sec:implementation} we describe our implementation and experimental results. We present the related work in section~\ref{sec:related}, we discuss possible extensions and future works in \autoref{sec:extensions}, and then, in \autoref{sec:conclusion} we draw the conclusions.

\section{Worst-Case Execution Time}
\label{sec:wcet}

\newcommand{\uAA}{Micro-architectural Analysis}
\tikzstyle{data}=[fill=black!3]
\tikzstyle{phase}=[fill=blue!7]
\tikzstyle{WCETall}  =[draw, text centered, text width=20mm, rounded corners=3pt]
\tikzstyle{WCETdata} =[WCETall, drop shadow, ellipse, inner sep=1pt, data]
\tikzstyle{WCETphase}=[WCETall, drop shadow, rectangle, phase]
\tikzstyle{legend}   =[rectangle, draw, rounded corners=1pt, text width=20mm]

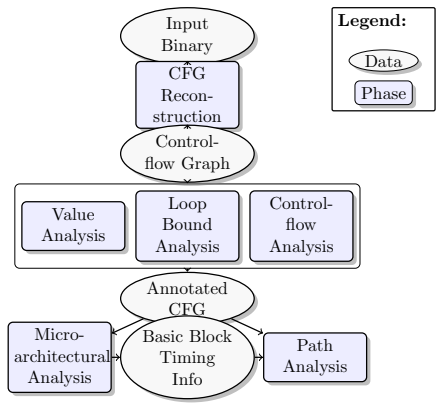
\begin{figure}\begin{center}
\scalebox{.6}{\begin{tikzpicture}[node distance=13mm, thick]
\node[WCETdata]  (BIN)                               {Input Binary};
\node[WCETphase] (CFGR) [below of=BIN]               {CFG Reconstruction};
\node[WCETdata]  (CFG)  [below of=CFGR]              {Control-flow Graph};
\node[WCETphase] (LBA)  [below of=CFG, yshift=-3mm]  {Loop Bound Analysis};
\node[WCETphase] (VA)   [left of =LBA, xshift=-12mm] {Value Analysis};
\node[WCETphase] (IPA)  [right of=LBA, xshift=+12mm] {Control-flow Analysis};
\node[WCETdata]  (ACFG) [below of=LBA, yshift=-3mm]  {Annotated CFG};
\node[WCETdata]  (BBT)  [below of=ACFG]              {Basic Block Timing Info};
\node[WCETphase] (uAA)  [left of =BBT, xshift=-15mm] {\uAA};
\node[WCETphase] (WPA)  [right of=BBT, xshift=+15mm] {Path\\Analysis};

\begin{pgfonlayer}{background}
    \node[WCETall, inner sep=4pt, fit=(VA) (LBA) (IPA), fill opacity=0.2] (MA) {};
\end{pgfonlayer}

\draw[->] (BIN)  -- (CFGR);
\draw[->] (CFGR) -- (CFG);
\draw[->] (CFG)  -- (MA);
\draw[->] (MA)   -- (ACFG);
\draw[->] (ACFG) -- (uAA);
\draw[->] (ACFG) -- (WPA);
\draw[->] (uAA)  -- (BBT);
\draw[->] (BBT)  -- (WPA);

\node[legend, anchor=north east, xshift=+15mm] (Legend) at (current bounding box.north east) {%
\textbf{Legend:}\\[0.5em]
\begin{center}
\begin{tikzpicture}[start chain=going below, node distance=2mm]
    \node[on chain, WCETdata, text width=10mm]  {Data};
    \node[on chain, WCETphase, text width=10mm] {Phase};
\end{tikzpicture}
\end{center}
};
\end{tikzpicture}}
\end{center}
\caption{WCET analysis workflow}
\label{wcet}
\end{figure}

Let us first summarize the classical approach to static timing analysis (for more detail, read e.g.~\parencite{Theiling_et_al_2000,wilhelm-et-TECS08}).
Figure~\ref{wcet} shows the general timing analysis workflow used in a large part of WCET tools including industrial ones such as AiT%
\footnote{\url{http://www.absint.com/ait/}}
or academic ones such as OTAWA%
\footnote{\url{http://www.otawa.fr}}%
~\parencite{DBLP:conf/seus/BallabrigaCRS10} or
\textsc{Chronos}%
\footnote{\url{http://www.comp.nus.edu.sg/~rpembed/chronos/}}%
~\parencite{Chronos_Li200756}.
For the sake of simplicity, we shall restrict ourselves to mono-processor platforms with no bus-master devices except for the CPU.

The analysis considers the object code.
The control flow graph is first reconstructed from the binary.
Then, a \emph{value analysis} (e.g. abstract interpretation for interval analysis) extracts memory addresses, loop bounds and simple infeasible paths \parencite{gustafsson-et-RTSS06};
such an analysis may be performed on the binary or the source files (in the latter case, it is necessary to trace object code and low-level variables to the source code, perhaps using the debugging information provided by the compiler).
This semantic and addressing information help the micro-architectural analysis,
which bounds the execution time of basic blocks taking into account the whole architecture of the platform (pipeline, caches, buses,...)\parencite{DBLP:conf/emsoft/EngblomJ02,DBLP:phd/de/Reineke2009}.
The most popular method to derive this architecture analysis is abstract interpretation with specific abstract domains.
For instance, a pipeline abstraction represents sets of detailed pipeline states, including values for registers or buffers \parencite{DBLP:conf/emsoft/EngblomJ02}; while a cache abstraction typically tracks which value may or must be in each cache line \parencite{DBLP:phd/de/Reineke2009}.

The last step of the analysis uses the basic block execution time and the semantic information to derive the WCET, usually, in the ``implicit path enumeration technique'' (IPET) approach, as the solution of an integer linear program (ILP)~\parencite{li-et-CADICS97}.
The ILP variables represent the execution counts (along a given trace) of each basic block in the program. The ILP constraints describe the structure of the control flow graph (e.g. the number of times a given block is entered equals the number of times it is exited), as well as maximal iteration counts for loops, obtained by value analysis or provided by the user. Finally, the execution time to be maximized is the sum of the basic blocks weighted by their local worst-case execution time computed by the microarchitectural analysis.

The obtained worst-case path may however be \emph{infeasible} semantically,
for instance, if a condition tests $x<10$ and later the unmodified value of $x$ is again tested in a condition $x>20$ along that path.
This is because the ILP represents mostly syntactic information from the control-flow graph.
This weakness has long been recognized within the WCET community, which has devised schemes for eliminating infeasible worst-case paths, for instance, by modifying the control-flow graph before the architecture analysis~\parencite{negi-et-WCET04}, or by adding ILP constraints~\parencite{healy-et-TSE02,gustafsson-et-RTSS06}.
Infeasible paths are found via pattern matching of conditions~\parencite{healy-et-TSE02} or applying abstract execution~\parencite{gustafsson-et-RTSS06}; these methods focus on paths made infeasible by numeric constraints.
These approaches are limited by the expressiveness of ILP constraints as used in IPET: they consider only ``conflict conditions'' (exclusive conditional statements: ``if condition $a$ is true then condition $b$ must be false'').

On a loop-free program, the ILP approach is equivalent to finding the longest path in the control-flow graph, weighted according to the local WCET of the basic blocks.
Yet, again, this syntactic longest path may be infeasible.
Instead of piecemeal elimination of infeasible paths, we propose encoding the set of feasible paths into an SMT formula, as done in bounded model-checking; the success of SMT-solving is based on the ability of SMT solvers to exclude whole groups of spurious solutions by learning lemmas.

Loop-free programs without recursion may seem a very restricted class, but in safety-critical control systems, it is common that the program consists in one big infinite control loop whose body must verify a WCET constraint, and this body itself does not contain loops, or only loops with small static bounds (say, for retrieving a value from an interpolation table of known static size), which can be unrolled.
Such programs typically eschew more complicated algorithms, if only because arguing for their termination or functional correctness would be onerous with respect to the stringent requirements imposed by the authorities. Complicated or dynamic data structures are usually avoided~\parencite[ch. II]{li-et-CADICS97}.
This is the class of programs targeted by e.g. the Astr\'ee static analyzer~\parencite{DBLP:conf/esop/CousotCFMMMR05}.

Our approach replaces the path analysis by ILP (and possibly refinement for infeasible paths) by optimization modulo theory. The control-flow extraction and micro-architectural analysis are left unchanged, and one may thus use existing WCET tools. In this paper we consider a simple architecture (ARMv7), though we plan to look into more complicated ones and address, for example, persistency analyses for caches, like in~\parencite{DBLP:conf/rtas/HuynhJR11}.

\section{Using Bounded Model Checking to Measure Worst-Case Execution Time}
\label{sec:bmc}

\begin{figure}
\begin{minipage}{0.3\textwidth}
\begin{lstlisting}
/* S */
if (b) {
  x = x + 2; /* C */
} else {
  x = x + 3; /* D */
}
assert(x >= 10);
\end{lstlisting}
\end{minipage}
\hfill
\begin{minipage}{\columnwidth}

First-order encoding:\\
$\left((b \land x_2 = x_1+2) \lor (\neg b \land x_2 = x_1+3)\right)
  \land x_2 \geq 10$
\medskip

Or, if the logic language comprises the ``if then else'' operator:
$\ite(b, x_1+2,x_1+3) \geq 10$
\medskip

If one wants to record the execution trace finely:

$(C \Leftrightarrow S \land b) \land (D \Leftrightarrow S \land \neg b)
\land\\(C \Rightarrow x_2 = x_1 + 2) \land (D \Rightarrow x_2 = x_1 + 3)
\land x_2 \geq 10
$
\end{minipage}

\caption{Encoding of a simple program into a first-order logic formula}
\label{fig:encoding}
\end{figure}

\emph{Bounded model checking} is an approach for finding software bugs, where traces of length at most $n$ are exhaustively explored.
In most current approaches, the set of feasible traces of length $n$ is defined using a first-order logic formula, where, roughly speaking, arithmetic constraints correspond to tests and assignments, control flow is encoded using Booleans, and disjunctions correspond to multiple control edges.
The source program may be a high-level language, an intermediate code (e.g. Java bytecode, LLVM bitcode \parencite{DBLP:conf/cgo/LattnerA04,DBLP:journals/entcs/HenryMM12}, Common Intermediate Language\dots) or even, with some added difficulty, binary executable code~\parencite{Chaki_Ivers_ISSE2010}.

The first step is to unroll all loops up to statically determined bounds.
Program variables and registers are then mapped to formula variables (implicitly existentially quantified). In an imperative language, but not in first-order logic, the same variable may be assigned several times: therefore, as in compilation to \emph{static single assignment} (SSA) form, different names have to be introduced for the same program variable, one for each update and others for variables whose value differs according to where control flows from (\autoref{fig:encoding}).
If the source program uses arrays or pointers to memory, the formula may need to refer not only to scalar variables, but also to \emph{uninterpreted functions} and \emph{functional arrays}~\parencite{Kroening_Strichman}.
Modern SMT-solvers support these datatypes and others suitable for the analysis of low-level programs, such as bit-vectors (fixed-width binary arithmetic).
If constructs occur in the source program that cannot be translated exactly into the target logic (e.g. the program has nonlinear arithmetic but the logic does not), they may be safely over-approximated by nondeterministic choice.
Details on ``conventional'' first-order encodings for program traces are given in the literature on bounded model checking \parencite{DBLP:journals/tse/CordeiroFM12} and are beyond the scope of this article.

Let us now see how to encode a WCET problem into SMT.
In a simple model (which can be made more complex and realistic, see \autoref{sec:extensions}), each program block $i$ has a fixed execution time $t_i \in \NN$, and the total execution time $T$ is the sum of the execution times of the blocks encountered in the trace. This execution time can be incorporated into a ``conventional'' encoding for program traces in two ways:

\begin{description}
\item[Sum encoding] If Booleans $\chi_i \in \{0,1\}$ record which blocks $i$ were reached by the execution trace $\tau$, then
\begin{equation}\label{eqn:sum_encoding}
T(\tau) = \left ( \sum_{i \mid \chi_i=\true} t_i \right)
        = \left (\sum_i \chi_i t_i \right)
\end{equation}
\item[Counter encoding]
Alternatively, the program may be modified by adding a time counter as an ordinary variable, which is incremented in each block.
The resulting program then undergoes the ``conventional'' encoding: the final value of the counter is the execution time.
\end{description}

An alternative is to attach a cost to transitions instead of program blocks. The sum encoding is then done similarly, with Booleans $\chi_{i,j} \in \{0,1\}$ recording which of the transitions have been taken by an execution trace $\tau$.
\begin{equation}\label{eqn:sum_encoding_edges}
T(\tau) = \left( \sum_{(i,j) \mid \chi_{i,j}=\true} t_{i,j} \right)
        = \left( \sum_{(i,j)} \chi_{i,j} t_{i,j} \right)
\end{equation}

The problem is now how to determine the WCET $\beta = \max T(\tau)$. An obvious approach is binary search~\parencite{Sebastiani_Tomasi_IJCAR12}, maintaining an interval $[l,h]$ containing $\beta$: take a middle point $m:=\lceil\frac{l+h}{2}\rceil$, test whether there exists a trace $\tau$ such that $T(\tau) \geq m$; if so, then set $l:=m$ (or set $l:=T(\tau)$, if available) and restart, else set $h:=m-1$ and restart; stop when the integer interval $[l,h]$ is reduced to a singleton. $l$ and $h$ may be respectively initialized to zero and a safe upper bound on worst-case execution time, for instance one obtained by a simple ``longest path in the acyclic graph'' algorithm.

\section{Adding Cuts}
\label{sec:cuts}
Experiments with both sum encoding and counter encoding applied to the ``conventional'' encoding of programs into SMT were disappointing: the SMT solver was taking far too much time. In particular, the last step of computing WCET, that is, running the SMT-solver in order to disprove the existence of traces longer than the computed WCET, was agonizingly slow even for very small programs.
In \autoref{sec:intractable} we shall see how this is inherent to how SMT-solvers based on DPLL(T) --- that is, all current production-grade SMT-solvers --- handle the kind of formulas generated from WCET constraints; but let us first see how we worked around this problem so as to make WCET computations tractable.

\subsection{Rationale}
A key insight is that the SMT-solver, applied to such a naive encoding, explores a very large number of combinations of branches (exponential with respect to the number of tests), thus a very large number of partial traces $\tau_1, \dots, \tau_n$, even though the execution time of these partial traces is insufficient to change the overall WCET (\autoref{sec:intractable} will explain this insight in more detail, both theoretically and experimentally).

\begin{figure}
\begin{center}
\scalebox{0.9}{


\tikzstyle{decision} = [diamond, draw, inner sep=0.3ex, every text node part/.style={align=center}, fill=red!7]
\tikzstyle{block} = [rectangle, draw, rounded corners, minimum height=3em, minimum width=4em, every text node part/.style={align=center}, fill=blue!7]
\tikzstyle{line} = [draw, -latex']

\begin{tikzpicture}[node distance = 1.7cm, auto, scale=0.7,every
	node/.style={scale=0.7}]
\node[block] (init) {block~1\\(start)};
\node[decision, below of=init] (dec1) {Decision 1};
\node[block, below right of=dec1] (blk2) {block~2};
\node[decision, below of=blk2] (dec2) {Decision 2};
\node[block, below left of=dec2] (blk3) {block~3};
\node[block, below right of=dec2] (blk4) {block~4};
\node[block, below right of=blk3] (blk5) {block~5};
\node[block, below left of=dec1] (blk6) {block~6};
\node[block, below left of=blk5] (blk7) {block~7\\(end)};
\path[line] (init)--(dec1);
\path[line] (dec1)--(blk2);
\path[line] (dec1)--(blk6);
\path[line] (blk2)--(dec2);
\path[line] (dec2)--(blk3);
\path[line] (dec2)--(blk4);
\path[line] (blk3)--(blk5);
\path[line] (blk4)--(blk5);
\path[line] (blk5)--(blk7);
\path[line] (blk6) to[out=250] (blk7);
\node[draw, dotted, fit=(dec2) (blk3) (blk4) (blk5)] (portion1) {};
\node[draw, dotted, fit=(dec1) (portion1) (blk6) (blk7)] {};
\node[above right of=portion1, xshift=0.3cm, yshift = 0.3cm] (p1) {\Large $P_2$};
\node[above right of=portion1, xshift=0.3cm, yshift = 3.2cm] (p1) {\Large $P_1$};

\end{tikzpicture}
}
\end{center}
\caption{Two portions $P_1$ and $P_2$ of a program obtained as the range between a node with several incoming edges and its immediate dominator}
\label{fig:portions}
\end{figure}
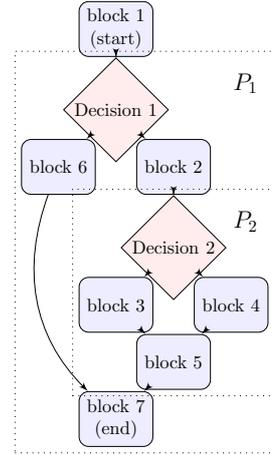

Consider the control-flow graph in \autoref{fig:portions}; let $t_1,\dots,t_7$ be the WCET of blocks $1\dots7$ established by microarchitectural analysis (for the sake of simplicity, we neglect the time taken for decisions).
Assume we have already found a path from start to end going through block~6, taking $\beta$ time units;
also assume that $t_1 + t_2 + \max(t_3, t_4) + t_5 + t_7 \leq \beta$.
Then it is useless for the SMT-solver to search for paths going through decision~2, because none of them can have execution time longer than $\beta$;
yet that is what happens if using a naive encoding with all current production SMT-solvers (see \autoref{sec:intractable}).
If instead of 1 decision we have $42$, then the solver may explore $2^{42}$ paths even though there is a simple reason why none of them will increase the WCET.

Our idea is simple: to the original SMT formula (from ``counter encoding'' or ``sum encoding''), conjoin constraints expressing that the total execution time of some portions of the program is less than some upper bound (depending on the portion). This upper bound acts as an ``abstraction'' or ``summary'' of the portion (e.g. here we say that the time taken in $P_2$ is at most $ \max(t_3, t_4) + t_5$), and the hope is that this summary is sufficient for the SMT-solver in many cases. There remain two problems: how to select such portions, and how to compute this upper bound.

Note that these extra constraints are implied by the original formula, and thus that conjoining them to it does not change the solution set or the WCET obtained, but only the execution time of the analysis. Such constraints are often called ``cuts'' in operation research, thus our terminology.

\subsection{Selecting portions}

The choice of a portion of code to summarize follows source-level criteria: for instance, a procedure, a block, a macro expansion.
If operating on a control-flow graph, a candidate portion can be between a node with several incoming edges and its \emph{immediate dominator}, if there is non trivial control flow between the two (Fig.~\ref{fig:portions}).%
\footnote{A \emph{dominator} $D$ of a block $B$ is a block such that any path reaching $B$ must go through $D$. The \emph{immediate dominator} of a block $B$ is the unique $I \neq B$ dominator of $B$ such that $I$ does not dominate any other dominator $D \neq B$ of $B$. For instance, the immediate dominator of the end of a cascade of if-then-else statements is the beginning of the cascade.}
On structured languages, this means that we add one constraint for the total timing of every ``if-then-else'' or ``switch'' statement (recall that loops are unrolled, if needed into a cascade of ``if-then-else'').
This is the approach that we followed in our experimental evaluation (\autoref{sec:implementation}).

Let us however remark that these portions of code need not be contiguous: with the sum encoding, it is straightforward to encode the fact that the total time of a number of instruction blocks is less than a bound, even though these instructions blocks are distributed throughout the code.
This is also possible, but less easy, with the counter encoding (one has to encode an upper bound on the sum of differences between starting and ending times over all contiguous subsets of the portion).
This means that it is possible to consider portions that are semantically, but not syntactically related.
For instance, one can consider for each Boolean, or other variable used in a test, a kind of ``slice'' of the program that is directly affected by this variable (e.g. all contents of if-then-elses testing on this variable) and compute an upper bound for the total execution time of this slice --- in the example in the introduction where the execution of two portions A and B depend on a variable \lstinline|clock|, we could compute an upper bound on the total time of the program sliced with respect to \lstinline|clock|, that only contains the portions A and B.
Implementing this ``slicing'' approach is part of our future work.

\subsection{Obtaining upper bounds on the WCET of portions}

Let us now consider the problem of, given a portion, computing an upper bound on its WCET.
In the case of a contiguous portion, an upper bound may be obtained by a simple syntactic analysis: the longest syntactic path is used as a bound (even though it might be unfeasible).
This approach may be extended to non-contiguous portions. Let us denote by $P$ the portion. For each block $b$, let $t_b$ be the upper bound on the time spent in block $b$ (obtained from microarchitectural analysis), and let $w_b$ be an unknown denoting the worst time spent inside $P$ in paths from the start of the program to the beginning of~$b$. If $b_1,\dots,b_k$ are the predecessors of $b$, then $w_b = \max(w_{b_1}+t_{b_1}.\chi_P(b_1), \dots, w_{b_k}+t_{b_k}.\chi_P(b_k))$ where $\chi_P(x)$ is $1$ if $x \in P$, $0$ otherwise.
This system of equations can be easily solved in (quasi) linear time by considering the $w_b$ in a topological order of the blocks (recall that we consider loop-free programs).
Another approach would be to recursively call the complete WCET procedure on the program portion, and use its output as a bound. 

The simpler approach described above gave excellent results in most benchmarks, and we had to refine the cuts with the SMT-based procedure for only one benchmark (see \autoref{sec:implementation}).
\subsection{Example}
\label{example}
Let us now see a short, but complete example, extracted from a control program composed of an initialization phase followed by an infinite loop clocked at a precise frequency. The goal of the analysis is to show that the WCET of the loop body never exceeds the clocking period. For the sake of brevity, we consider only a very short extract of the control program, implementing a ``rate limiter''; in the real program its input is the result of previous computation steps, but here we consider that the input is nondeterministic within $[-10000,+10000]$. The code run at every clock tick is:

\begin{lstlisting}[tabsize=4]
// returns a value between min and max
extern int input(int min, int max);
void rate_limiter_step() {
	int x_old = input(-10000,10000);
	int x = input(-10000,10000);
	if (x > x_old+10)
		x = x_old+10;
	if (x < x_old-10)
		x = x_old-10;
	x_old = x;
}

\end{lstlisting}

This program is compiled to LLVM bitcode,%
\footnote{LLVM (\url{http://www.llvm.org/}) \parencite{DBLP:conf/cgo/LattnerA04} is a compilation framework with a standardized intermediate representation (bitcode), into which one can compile with a variety of compilers including GCC (C, C++, Ada\dots) and \textsc{Clang} (C, C++).}
then bitcode-level optimizations are applied, resulting in a LLVM control-flow graph (\autoref{fig:LLVM_cfg_formula} left). From this graph we generate a first-order formula including cuts (\autoref{fig:LLVM_cfg_formula} right). Its models describe execution traces along with the corresponding execution time \emph{cost} given by the ``sum encoding''.
Here, costs are attached to the transitions between each pairs of blocks. These costs are supposed to be given. \Cref{sec:realistic_timing} will describe in full details how we use the \textsc{OTAWA} tool to derive such precise costs for each transitions.

The SMT encoding of the program semantics (\autoref{fig:LLVM_cfg_formula} right) is relatively simple since the bitcode has an SSA form:
The $ite(b,x,y)$ construct is an \emph{if-then-else} statement and is equal to $x$ if $b$ is true, otherwise is equal to $y$. 
In our encoding, SMT variables starting with letter $x$ refer to the LLVM SSA-variables, there is one Boolean $b\_i$ for each LLVM BasicBlock, and one Boolean $t\_i\_j$ for each transition. Each transition $t\_i\_j$ have a cost $c\_i\_j$ given by \textsc{OTAWA}.
For instance, the block \texttt{entry} is given the Boolean $b\_0$, the block \texttt{if.then} is given the Boolean $b\_1$, and the transition from \texttt{entry} to \texttt{if.then} is given the Boolean $t\_0\_1$ and has a cost of $15$ clock cycles.
The cuts are derived as follows: \texttt{if.end} has several incoming transitions and its immediate dominator is \texttt{entry}. The longest syntactic path between these two blocks is equal to 21. The cut will then be $c\_0\_1 + c\_1\_2 + c\_0\_2 \leq 21$. There is a similar cut for the portion between \texttt{if.end} and \texttt{if.end6}. Finally, we can also add the constraint $cost \leq 43$ since it is the cost of the longest syntactic path.
While this longest syntactic path has cost $43$ (it goes both through \texttt{if.then} and \texttt{if.then4}), our SMT-based approach shows there is no semantically feasible path longer than $36$ clock cycles.

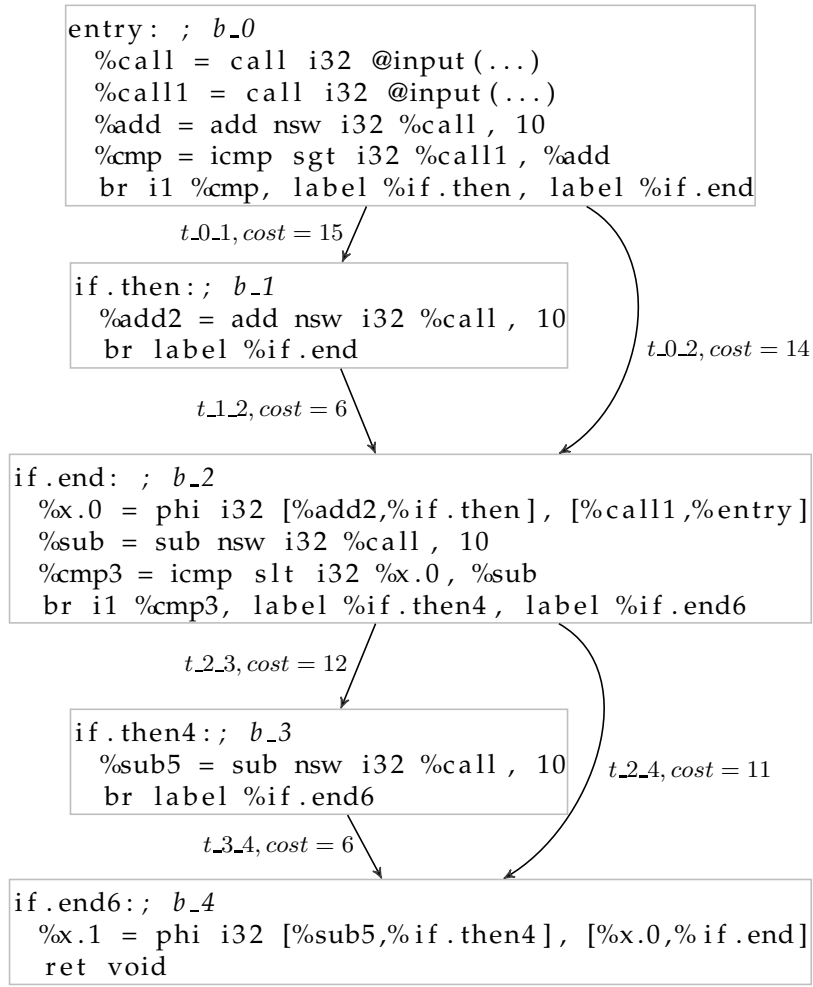
\begin{figure*}
\begin{center}
\usetikzlibrary{snakes,arrows,shapes,backgrounds,shadows,automata,patterns}
\tikzstyle{state}=[rectangle,draw=black!25,minimum size=13pt,inner sep=0pt]
\tikzstyle{transition}=[rectangle,semithick,draw=black!75,
  			  minimum size=4mm]

\begin{tikzpicture}[->,>=stealth',auto,node distance=1.7cm,
                    semithick,font=\footnotesize]

	\node[state] (n0) {
		\begin{llvm}
entry: ; b_0
  %call = call i32 @input(...)
  %call1 = call i32 @input(...)
  %add = add nsw i32 %call, 10
  %cmp = icmp sgt i32 %call1, %add
  br i1 %cmp, label %if.then, label %if.end
		\end{llvm}
	};
	\node[state] (n1) [below left of=n0,yshift=-45pt] {
		\begin{llvm}
if.then:; b_1
  %add2 = add nsw i32 %call, 10
  br label %if.end			
		\end{llvm}
	};
	\node[state] (n2) [below right of=n1,yshift=-50pt] {
		\begin{llvm}
if.end: ; b_2
  %x.0 = phi i32 [%add2,%if.then], [%call1,%entry]
  %sub = sub nsw i32 %call, 10
  %cmp3 = icmp slt i32 %x.0, %sub
  br i1 %cmp3, label %if.then4, label %if.end6	
		\end{llvm}
	};
	\node[state] (n3) [below left of=n2,yshift=-50pt] {
		\begin{llvm}
if.then4:; b_3
  %sub5 = sub nsw i32 %call, 10
  br label %if.end6	
		\end{llvm}
	};
	\node[state] (n4) [below right of=n3,yshift=-30pt] {
		\begin{llvm}
if.end6:; b_4
  %x.1 = phi i32 [%sub5,%if.then4], [%x.0,%if.end]
  ret void	
		\end{llvm}
	};

  \path [transition] 
  (n0) edge              node [left] {$t\_0\_1, cost=15$} (n1);
  \path [transition] 
		(n0) edge [out=-30,in=30] node {$t\_0\_2, cost=14$} (n2);
  \path [transition] 
  (n1) edge              node [left] {$t\_1\_2, cost=6$} (n2);
  \path [transition] 
  (n2) edge              node [left] {$t\_2\_3, cost=12$} (n3);
  \path [transition] 
  (n2) edge [out=-30,in=30]    node {$t\_2\_4, cost=11$} (n4);
  \path [transition] 
  (n3) edge              node  [left] {$t\_3\_4, cost=6$} (n4);

\end{tikzpicture}
\end{center}
\caption{LLVM control-flow graph of the rate\_limiter\_step function. 
} 
\label{fig:LLVM_cfg}
\end{figure*}

\begin{figure*}
	\small
$$
\begin{blockarray}{ll}
  & -10000 \leq x\_call	\leq 10000 \\
 \wedge & -10000 \leq x\_call1\leq 10000 \\
 \wedge & x\_add = (x\_call + 10) \\
 \wedge & t\_0\_1 = (b\_0 \wedge (x\_call1 > x\_add)) \\
 \wedge & t\_0\_2 = (b\_0 \wedge \neg(x\_call1 > x\_add)) \\
 \wedge & b\_1 = t\_0\_1 \\
 \wedge & x\_add2 = (x\_call + 10) \\
 \wedge & t\_1\_2 = b\_1 \\
 \wedge & b\_2 = (t\_0\_2 \vee t\_1\_2) \\
 \wedge & b\_2 \Rightarrow (x\_x.0 = ite(t\_1\_2,x\_add2,x\_call1)) \\
 \wedge & x\_sub = (x\_call - 10) \\
 \wedge & t\_2\_3 = (b\_2 \wedge (x\_x.0 < x\_sub)) \\
 \wedge & t\_2\_4 = (b\_2 \wedge \neg(x\_x.0 < x\_sub)) \\
 \wedge & b\_3 = t\_2\_3 \\
 \wedge & x\_sub5 = (x\_call - 10) \\
 \wedge & t\_3\_4 = b\_3 \\
 \wedge & b\_4 = (t\_2\_4 \vee t\_3\_4) \\
 \wedge & b\_4 \Rightarrow (x\_x.1 = ite(t\_3\_4,x\_sub5,x\_x.0)) \\
  & \\
  \wedge & b\_0 = b\_4 = true \textnormal{~;~ search for a trace from entry to if.end6}\\
 \begin{block}{\Left{\rotatebox{90}{timing}}{\{}ll}
	 \wedge & c\_0\_1 = ite(t\_0\_1,15,0) \textnormal{~ ; ~ $t\_0\_1$ has cost 15 if taken, else 0} \\
 \wedge & c\_0\_2 = ite(t\_0\_2,14,0)\\
 \wedge & c\_1\_2 = ite(t\_1\_2,6,0)\\
 \wedge & c\_2\_3 = ite(t\_2\_3,12,0)\\
 \wedge & c\_2\_4 = ite(t\_2\_4,11,0)\\
 \wedge & c\_3\_4 = ite(t\_3\_4,6,0)\\
 \wedge & cost = (c\_0\_1 + c\_0\_2 + c\_1\_2 + c\_2\_3 + c\_2\_4 + c\_3\_4) \\
 \end{block}
 \begin{block}{\Left{\rotatebox{90}{cuts}}{\{}ll}
	 \wedge & (c\_0\_1 + c\_1\_2 + c\_0\_2) \leq 21 \textnormal{~ ;~ between entry and if.end}\\
 \wedge & (c\_3\_4 + c\_2\_4 + c\_2\_3) \leq 22 \textnormal{~ ;~ between if.end and if.end6}\\
 \wedge & cost \leq 43\\
 \end{block}
\end{blockarray}
$$
\caption{Encoding of the rate\_limiter\_step function (control-flow graph on Fig.~\ref{fig:LLVM_cfg}) as an SMT formula with cuts.} 
\label{fig:LLVM_cfg_formula}
\end{figure*}

\subsection{Relationship with Craig interpolants}
\label{sec:interpolants}
A \emph{Craig interpolant} for an unsatisfiable conjunction $F_1 \land F_2$ is a formula $I$ such that $F_1 \Rightarrow I$ and $I \land F_2$ is unsatisfiable, whose free variables are included in the intersection of those of $F_1$ and $F_2$.

In the case of a program $A ; B$ consisting of two portions $A$ and $B$ executed in sequence, the usual way of encoding the program yields $\phi_A \land \phi_B$ where $\phi_A$ and $\phi_B$ are, respectively, the encodings of $A$ and~$B$.
The free variables of this formula are the inputs $i_1,\dots,i_m$ and outputs $o_1,\dots,o_n$ of the program, as well as all temporaries and local variables.
Let $t_1,\dots,t_p$ be the variables live at the edge from $A$ to $B$; then the input-output relationship of the program, with free variables $i_1,\dots,i_m,o_1,\dots,o_n$ is $F$~:
\begin{equation*}
\exists t_1,\dots,t_p (\exists \dots \phi_A) \land (\exists \dots \phi_B)
\end{equation*}

Let us now assume additionally that $o_1$ is the final time and $t_1$ is the time when control flow from $A$ to $B$ (counter encoding). The SMT formulas used in our optimization process are of the form $F \land t_1 \geq \beta$.
The cut for portion $A$ is of the form $t_1 \leq \beta_A$, that for portion $B$ of the form $o_1 - t_1 \leq \beta_B$. Then, if the cut for portion $A$ is used to prove that $F \land t_1 \geq \beta$ is unsatisfiable, then this cut is a Craig interpolant for the unsatisfiable formula $(\phi_A) \land (\phi_B \land t_1 \geq \beta)$ (similarly, if the cut for portion $B$ is used, then it is an interpolant for $\phi_B \land (\phi_A \land t_1 \geq \beta)$.
Our approach may thus be understood as preventively computing possible Craig interpolants so as to speed up solving.
The same intuition applies to the sum encoding (up to the creation of supplementary variables).

\section{Intractability: Diamond Formulas}
\label{sec:intractable}
Let us now explain why the formulas without cuts result in unacceptable execution times in the SMT-solvers.

\begin{figure}
\noindent\hspace*{-1.2ex}
\centering
\input{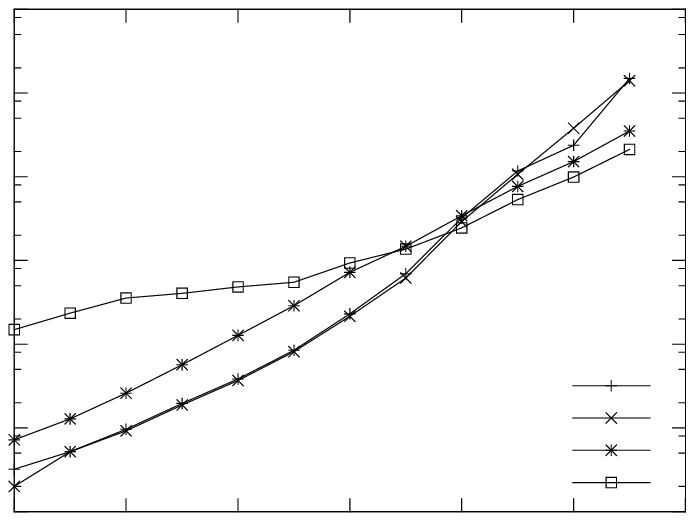}
\caption{Intractability of diamond formulas obtained from timing analysis of a family of programs with very simple functional semantics.
Execution times of various state-of-the-art SMT-solvers on \autoref{constraint}, for $m=5n$ (the hardest), showing exponential behavior in the formula size~$n$.
The CPU is a 2~GHz Intel Core~2 Q8400.}
\label{fig:intractability}  
\end{figure}



Consider a program consisting in a sequence of $n$ fragments where the $i$-th
fragment is of the form:
\begin{lstlisting}
if ($b_i$) { /* block of cost $x_i$ */
  /* time cost 2, not changing $b_i$ */
} else {
  /* time cost 3, not changing $b_i$ */
}
if ($b_i$) { /* block of cost $y_i$ */
  /* time cost 3 */
} else {
  /* time cost 2 */
}
\end{lstlisting}
The $(b_i)_{1 \leq i \leq n}$ are Booleans.
A human observer easily concludes that the worst-case execution time is~$5n$, by analyzing each fragment separately.

Using the ``sum encoding'', the timing analysis is expressed as
\begin{multline}
T = \max \bigg\{ \sum_{i=1}^n x_i + y_i ~\bigg|~
\bigwedge_{i=1}^n (x_i=\ite(b_i,2,3))\\ \land (y_i=\ite(b_i,3,2)) \bigg\}
\end{multline}

Given a bound $m$, an SMT-solver will have to solve for the unknowns
$(b_i), \allowbreak (x_i), \allowbreak (y_i)_{1 \leq i \leq n}$ the constraint
{\small
\begin{multline}\label{constraint}
\big((b_1 \land x_1=2 \land y_1=3) \lor (\neg b_1 \land x_1=3 \land y_1=2)\big) \land \dots\\
\big((b_n \land x_n=2 \land y_n=3) \lor (\neg b_n \land x_n=3 \land y_n=2)\big) \land \\
x_1 + y_1 + \dots + x_n + y_n \geq m
\end{multline}}
In the ``DPLL(T)'' approach (see e.g. \textcite{Kroening_Strichman} for an introduction), SMT is implemented as a combination of a SAT solver,%
\footnote{Almost all current SAT solvers implement variants of \emph{constraint-driven clause learning} (CDCL), a major improvement over DPLL (Davis, Putnam, Logemann, Loveland), thus the terminology. None of what we say here, however, is specific to CDCL: our remarks stay valid as long as the combination of propositional and theory reasoning proceeds by sending clauses constructed from the predicates syntactically present in the original formula to the propositional solver.}
which searches within a Boolean state space (here, amounting to $b_1,\dots,b_n \in \{0,1\}^n$, but in general arithmetic or other theory predicates are also taken into account) and a decision procedure for conjunctions of atomic formulas from a theory~T.%
\footnote{We leave out improvements such as \emph{theory propagation} for the sake of simplicity. See \textcite{Kroening_Strichman} for more details.}
Once $b_1,\dots,b_n$ have been picked, \autoref{constraint} simplifies to a conjunction
\begin{multline}\label{conj:diamond_cube}
x_1=\alpha_1 \land y_1=\beta_1 \land \allowbreak
\dots \allowbreak \land \allowbreak
x_n=\alpha_n \land y_n=\beta_n \allowbreak\\
\land \allowbreak
x_1 + y_1 + \dots + x_n + y_n \geq m
\end{multline}
where the $\alpha_i,\beta_i$ are constants in $\{2,3\}$ such that for each $i$,
$\alpha_i + \beta_i = 5$.
Such a formula is satisfiable if and only if $m \leq 5n$.

Assume now $m > 5n$. All combinations of $b_1,\dots,b_n$ lead to unsatisfiable constraints, thus \autoref{constraint} is unsatisfiable. Such an exhaustive exploration is equivalent to exploring $2^n$ paths in the control flow graph, computing the execution time for each and comparing it to the bound. Could an SMT-solver do better?
SMT-solvers, when exploring the Boolean state space, may detect that the current Boolean choices (say, $b_3 \land \neg b_5 \land b_7$) lead to an arithmetic contradiction, without picking a value for all the Booleans. The SMT-solver extracts a (possibly smaller) contradiction (say, $b_3 \land \neg b_5$), adds the negation of this contradiction to the Boolean constraints as a \emph{theory clause}, and restarts Boolean solving.
The hope is that there exist short contradictions that enable the SMT-solver to prune the Boolean search space.
Yet, in our case, there are no such short contradictions: if one leaves out \emph{any} of the conjuncts in \autoref{conj:diamond_cube}, the conjunction becomes satisfiable.
Note the asymmetry between proving satisfiability and unsatisfiability: for satisfiability, one can always hope that clever heuristics will lead to one solution, while for unsatisfiability, the prover has to close all branches in the search.

The difficulty of \autoref{constraint} or similar ``diamond formulas'' is well-known in SMT circles.
It boils down to the SMT-solver working exclusively with the predicates found in the original formulas, without deriving new useful ones such as $x_i+y_i \leq 5$.
All state-of-the-art solvers that we have tried have exponential running time in $n$ when solving \autoref{constraint} for $m = 5n$ (\autoref{fig:intractability})%
\footnote{A special version of MathSAT~5, which was kindly made available to us by the authors~\parencite{Sebastiani_Tomasi_IJCAR12}, implements the binary search approach internally. It suffers from the same exponential behavior as noted in the figure: in its last step, it has to prove that the maximum obtained truly is maximum.};
the difficulty increases exponentially as upper bound on the WCET to be proved becomes closer to the actual WCET.

There have been proposals of alternative approaches to DPLL(T), where one would
directly solve for the numeric values instead of solving for Booleans then
turning theory lemmas into Boolean constraints
\parencite{Cotton_PhD,DBLP:conf/formats/Cotton10,DBLP:conf/cav/McMillanKS09,Bjorner_et_al_LPAR2008,DBLP:conf/vmcai/MouraJ13};
but no production solver implements them.%
\footnote{Dejan Jovanovic was kind enough to experiment with some of our formulas in his experimental solver \parencite{DBLP:conf/vmcai/MouraJ13}, but the execution time was unacceptably high. We stress that this field of workable alternatives to DPLL(T) is still new and it is too early to draw conclusions.}
This is the reason why we turned to incorporating cuts into the encoding.

\section{Implementation and Experimental Results}
\label{sec:implementation}

We experimented our approach for computing the worst-case execution time on
benchmarks from several sources, referenced in \autoref{tab:benchmarkslist}.
\texttt{nsichneu} and \texttt{statemate} belong to the M\"alardalen WCET benchmarks set 
\parencite{Gustafsson:WCET2010:Benchmarks}\footnote{\url{http://www.mrtc.mdh.se/projects/wcet/benchmarks.html}}, being the largest of the set (w.r.t. code size).
\texttt{cruise-control} and \texttt{digital-stopwatch} are generated from \textsc{Scade}{\RegTM} designs.
\texttt{autopilot} and \texttt{fly-by-wire} come from the Papabench benchmark \parencite{DBLP:conf/wcet/NemerCSBM06} derived from the Paparazzi free software suite for piloting UAVs (\url{http://paparazzi.enac.fr/}).
\texttt{tdf} and \texttt{miniflight} are industrial avionic case-studies.

\subsection{Description of the Implementation}
We use the infrastructure of the \textsc{Pagai} static analyzer 
\parencite{DBLP:journals/entcs/HenryMM12}%
\footnote{\url{http://pagai.forge.imag.fr}}
to produce an SMT formula
corresponding to the semantics of a program expressed in LLVM bitcode.

A limitation is that, at present, \textsc{Pagai} considers that floating-point variables are real numbers and that integers are unbounded mathematical integers, as opposed to finite bit-vectors; certainly an industrial tool meant to provide sound bounds should have accurate semantics, but this limitation is irrelevant to our proof-of-concept (note how the bitvectors from functional semantics and the timing variables are fully separated --- their combination would therefore not pose a problem to any SMT-solver implementing a variant of the Nelson-Oppen combination of procedures~\parencite[ch.~10]{Kroening_Strichman}).

Using the LLVM optimization facilities, we first apply some standard transformation to the program (loop unrolling, function inlining, SSA) so as to obtain a single loop-free function; in a manner reminiscent of bounded model checking.
Once the SMT formula is constructed, we enrich it with an upper timing bound for each basic block.

Finally, we conjoin to our formula the cuts for the ``sum
encoding'', i.e., constraints of the form $\sum_{i\in S} c_i \leq B$, where the
$c_i$'s are the cost variables attached to the basic blocks.
There is one such ``cut'' for every basic block with several incoming edges:
the constraint expresses an upper bound on the total timing of the program portion comprised between the block and its immediate dominator (Fig.~\ref{fig:portions}).
The bound $B$ is the weight of the maximal path through the range, be it feasible or infeasible (a more expensive method is to call the WCET computation recursively on the range).

We use Z3 \parencite{DBLP:conf/tacas/MouraB08} as an SMT solver and a binary search strategy to maximize the \emph{cost} variable modulo SMT.

The encoding of program semantics into SMT may not be fully precise in some cases.
Whenever we cannot precisely translate a construct from the LLVM bitcode, we abstract it by nondeterministic choices into all the variables possibly written to by the construct (an operation referred to as \texttt{havoc} in certain systems); for instance, this is the case for loads from memory locations that we cannot trace to a specific variable. We relied on the LLVM \verb+mem2reg+ optimization phase to lift memory accesses into SSA (single static assignment) variables; all accesses that it could not lift were thus abstracted as nondeterministic choice. We realized that, due to being limited to local, stack-allocated variables, this phase missed some possible liftings, e.g. those of global variables. This resulted in the same variable from the program to be analyzed being considered as several unrelated nondeterministic loads from memory, thereby breaking dependencies between tests and preventing infeasible paths from being discarded. We thus implemented a supplemental lifting phase for global variables.
It is however possible that our analysis still misses infeasible paths because of badly abstracted constructs (for instance, arrays), and that further improvements could bring even better results (that is, upper bounds on the WCET that would be closer to the real WCET).

Furthermore, some paths are infeasible because of a global invariant of the control loop (e.g. some Booleans $a$ and $b$ activate mutually exclusive modes of operations, and $\neg a \lor \neg b$ is an invariant); we have not yet integrated such invariants, which could be obtained either by static analysis of the program, either by analysis of the high-level specification from which the program is extracted~\parencite{Asavoae_et_al-WCET2013}.

Our current implementation keeps inside the program the resulting formulas statements and variables that have no effect on control flow and thus on WCET. Better performance could probably be obtained by slicing away such irrelevant statements.

\subsection{Results with bitcode-based timing}
The problem addressed in this article is not architectural modeling and low-level timing analysis: we assume that worst-case timings for basic blocks are given by an external analysis.
Here we report on results with a simple timing basis: the time taken by a LLVM bitcode block is its number of instructions;
our goal here is to check whether improvements to WCET can be obtained by our analysis with reasonable computation costs, independently of the architecture.

\begin{table}\small
\centering
\begin{tabular}{|l|r|r|} \hline
	Benchmark name & LLVM \#lines & LLVM \#BB \\ \hline
statemate  	&  2885 	&  632  \\ 
nsichneu  	&  12453 	&  1374  \\ 
cruise-control  	&  234 	&  43  \\ 
digital-stopwatch  	&  1085 	&  188  \\ 
autopilot  	&  8805 	&  1191  \\ 
fly-by-wire  	&  5498 	&  609  \\ 
miniflight  	&  5860 	&  745  \\ 
tdf  	&  2689 	&  533  \\ 

	\hline
\end{tabular}
\caption{Table referencing the various benchmarks. LLVM \#lines is the number of lines in the LLVM bitcode, and LLVM \#BB is its number of Basic Blocks.}
\label{tab:benchmarkslist}
\end{table}

\begin{table*}\small\centering
\begin{tabular}{|l|r|r|r|r|r|r|} \hline
  & \multicolumn{3}{c|}{WCET bounds} & \multicolumn{2}{c|}{Analysis time (in seconds)} & \\
  Benchmark name & \multicolumn{1}{c|}{syntactic/OTAWA} & \multicolumn{1}{c|}{max-SMT}
  & \multicolumn{1}{c|}{diff} & \multicolumn{1}{c|}{with cuts} &
  \multicolumn{1}{c|}{without cuts} & \#cuts \\\hline \hline
  \multicolumn{7}{|c|}{Bitcode-based timings (in number of LLVM instructions)} \\ \hline
  statemate  	&  997 	& 951 	& 4.6\% 	& 118.3 	&  $+\infty$ &  143  \\ 
nsichneu  	&  9693 	& 5998 	& 38.1\% 	& 131.4 	&  $+\infty$ 	&  252  \\ 
cruise-control 	&  123 	& 121 	& 1.6\% 	& 0.1 	&  0.1	&  13  \\ 
digital-stopwatch  	&  332 	& 302 	& 9.0\% 	& 1.0 	&  35.5	&  53  \\ 
autopilot 	&  4198 	& 1271	& 69.7\% 	& 782.0 	&  $+\infty$	&  498  \\ 
fly-by-wire  	&  2932 	& 2792 	& 4.7\% 	& 7.6 	&  $+\infty$	&  163  \\ 
miniflight  	&  4015 	& 3428 	& 14.6\% 	& 35.8 	&  $+\infty$	&  251  \\ 
tdf  	&  1583 	& 1569 	& 0.8\% 	& 5.4 	&  343.8	&  254  \\ 

  \hline
  \multicolumn{7}{|c|}{Realistic timings (in cycles) for an ARMv7 architecture} \\ \hline
statemate  	&  3297 	& 3211 	& 2.6\% 	& 943.5 	&  $+\infty$ 	&  143  \\ 
nsichneu* (1 iteration)  	&   17242	& $<$13332** & 22.7\% 	& 3600** 	&  $+\infty$	& 378   \\ 
cruise-control 	&  881 	& 873	& 0.9\% 	& 0.1 	&  0.2	&  13  \\ 
digital-stopwatch  	&  1012 	& 954 	& 5.7\% 	& 0.6 	&  2104.2	&  53  \\ 
autopilot  	&  12663 	& 5734	& 54.7\% 	& 1808.8 	&  $+\infty$	&  498  \\ 
fly-by-wire  	&  6361 	& 5848 	& 8.0\% 	& 10.8 	&  $+\infty$	&  163  \\ 
miniflight   	&  17980 	& 14752 	& 18.0\% 	& 40.9 	&  $+\infty$	&  251  \\ 
tdf  	&  5789 	& 5727 	& 1.0\% 	& 13.0 	&  $+\infty$	&  254  \\ 

  \hline
\end{tabular}
\caption{\emph{max-SMT} is the upper bound on WCET reported by our analysis based on optimization modulo theory, while \emph{syntactic/OTAWA} is the execution time of longest syntactic path (provided by Otawa when using realistic timings). 
\emph{diff} is the improvement brought by our method. The analysis time for \emph{max-SMT} is reported with and without added cuts; $+\infty$ indicates timeout (1~hour). 
\emph{\#cuts} is the number of added cuts. In the second part, *) \texttt{nsichneu} has been simplified to one main-loop iteration (instead of 2), and has been computed with cuts refinement as described in \autoref{sec:realistic_timing}. **) Computation takes longer than 1 hour. A safe bound of 13332 is already known after this time.
}
\label{tab:benchmarks}
\end{table*}

As expected, the naive approach (without adding cuts to the formula) does not scale at all, and the computation has reached our timeout in all of our largest benchmarks. Once the cuts are conjoined to the formula, the WCET is computed considerably faster, with some benchmarks needing less than a minute while they timed out with the naive approach.

Our results (\autoref{tab:benchmarks}, first part) show that the use of bounded model checking by SMT solving improves
the precision of the computed upper bound on the worst-case execution time,
since the longest syntactic path is in most cases not feasible due to the semantics of the instructions.
As usual with WCET analyzes, it is difficult to estimate the absolute quality of the resulting bound, because the exact WCET is unknown (perhaps what we obtain is actually the WCET, perhaps it overestimates it somewhat).

On the \texttt{autopilot} software, our analysis reduces the WCET bound by 69.7\%. This software has multiple clock domains, statically scheduled by the \verb+periodic_task()+ function using switches and arithmetic constraints. Approaches that do not take functional semantics into account therefore consider activation patterns that cannot occur in the real system, leading to a huge overestimation compared to our semantic-sensitive approach.

\subsection{Results with realistic timing}
\label{sec:realistic_timing}

\tikzstyle{block} = [rectangle, draw, rounded corners, minimum height=3em, minimum width=4em, every text node part/.style={align=center}, fill=blue!7]
\tikzstyle{metablock} = [rectangle, draw, dotted, rounded corners, minimum height=3em, minimum width=4em, every text node part/.style={align=center}]
\tikzstyle{blanc} = [rectangle, fill=white, rounded corners, minimum height=3em, minimum width=5em,  every text node part/.style={align=left}]
\tikzstyle{line} = [draw, -latex']

The timing model used in the preceding subsection is not meant to be realistic. We therefore experimented with realistic timings for the basic blocks, obtained by the OTAWA tool~\parencite{DBLP:conf/seus/BallabrigaCRS10} for an ARMv7 architecture. The results are given in \autoref{tab:benchmarks} (second half).

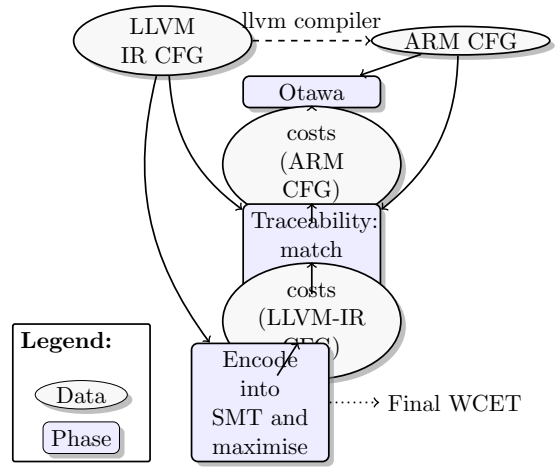
\begin{figure}[h]
	\centering
\scalebox{.8}{
\begin{tikzpicture}[node distance=12mm, thick]
\node[WCETdata]  (BC)                               {LLVM IR CFG};
\node[WCETdata]  (ARM)  [right of=BC,xshift=3.7cm]    {ARM CFG};
\node[WCETphase] (OTAWA) [below left of=ARM,xshift=-1.6cm]              {Otawa};
\node[WCETdata] (ARMCOSTS) [below of=OTAWA]              {costs (ARM CFG)};
\node[WCETphase] (TRAC) [below of=ARMCOSTS,yshift=-2mm]              {Traceability: match blocks};
\node[WCETdata] (BCCOSTS) [below of=TRAC]              {costs (LLVM-IR CFG)};
\node[WCETphase] (SMT) [below left of=BCCOSTS,yshift=-0.5cm]              {Encode into SMT and maximise};
\node (WCET) [right of=SMT,xshift=2cm]              {Final WCET};

\draw[->,dashed] (BC) to node[midway,above] {llvm compiler}
       (ARM);
\draw[->] (BC)  to [bend right=25] (TRAC);
\draw[->] (ARM)  to [bend left=25] (TRAC);
\draw[->] (ARM)  -- (OTAWA);
\draw[->] (OTAWA)  -- (ARMCOSTS);
\draw[->] (ARMCOSTS)  -- (TRAC);
\draw[->] (TRAC)  -- (BCCOSTS);
\draw[->] (BCCOSTS)  -- (SMT);
\draw[->] (BC) to [bend right=25]  (SMT);
\draw[->,dotted] (SMT)  -- (WCET);

\node[legend, anchor=south west, xshift=-10mm] (Legend) at (current bounding box.south west) {%
\textbf{Legend:}\\[0.5em]
\begin{center}
\begin{tikzpicture}[start chain=going below, node distance=2mm]
    \node[on chain, WCETdata, text width=10mm]  {Data};
    \node[on chain, WCETphase, text width=10mm] {Phase};
\end{tikzpicture}
\end{center}
};
\end{tikzpicture}
}
	\caption{General workflow for deriving timings using OTAWA.}
	\label{fig:flow}

\end{figure}

The difficulty here is that OTAWA considers the basic blocks occurring in binary code, while our analysis considers the basic blocks in the LLVM bitcode. 
The LLVM blocks are close to those in the binary code, but code generation slightly changes the block structure in some cases. The matching of binary code to LLVM bitcode is thus sometimes imperfect and we had to resort to one that safely overestimates the execution time. \autoref{fig:flow} gives an overview of the general workflow for deriving the appropriate costs of LLVM basic blocks.
The alternative would be to generate the SMT formulas not from LLVM bitcode, but directly from the binary code; unfortunately a reliable implementation needs to address a lot of open questions, and as such, it falls into our future plans.

While the \texttt{nsichneu} benchmark is fully handled by our approach when using bitcode-based timing, it is much harder when using the realistic metric. We had to improve our implementation in two ways:
\begin{enumerate*}
	\item We extract cuts for larger portions of the program: we take the portions from our previous cuts (between merge points and their immediate dominators) and derive new cuts by recursively grouping these portions by two. We then have cuts for one half, one quarter, etc. of the program.
	\item Instead of directly optimising the total cost variable of the program, we successively optimize the variables expressing the ``cuts'' (in order of portion size). This allows to strengthen the cuts with smaller upper bounds, and helps the analysis of the bigger portions.
\end{enumerate*}
In this benchmark, all the biggest paths are unfeasible because of inconsistent semantic constraints over the variables involved in the tests. Better cuts could be derived if we were not restricted to contiguous portions in the implementation. The computation time is around 6.5 hours to get the exact WCET (13298 cycles), but we could have stopped after one hour and get a correct upper bound of 13332 cycles, which is already very close to the final result.

\section{Related Work}
\label{sec:related}

The work closest to ours is from \textcite{Chu_Jaffar_EMSOFT2011}.
They perform symbolic execution on the program, thereby unrolling an exponentially-sized execution tree (each if-then-else construct doubles the number of branches).
This would be intolerably expensive if not for the very astute subsumption criterion used to fold some of the branches into others already computed.
More specifically, their symbolic execution generalizes each explored state $S$ to a first-order formula defining states from which the feasible paths are included in those starting from $S$;
these formula are obtained from \emph{Craig interpolants} extracted from the proofs of infeasibility.

In our approach, we also learn formula that block infeasible paths or paths that cannot lead to paths longer than the WCET obtained, in two ways: the SMT-solver learns blocking clauses by itself, and we feed ``cuts'' to it.
Let us now attempt to give a high-level view of the difference between our approach and theirs.
Symbolic execution \parencite{Cadar:2013:SES:2408776.2408795} (in depth-first fashion) can be simulated by SMT-solving by having the SMT-solver select decision literals \parencite{Kroening_Strichman} in the order of execution of the program encoded into the formula;
in contrast, general bounded model checking by SMT-solving will assert predicates in an arbitrary order, which may be preferrable in some cases (e.g. if $x \leq 0$ is asserted early in the program and $x + y \geq 0$ very late, after multiple if-then-elses, it is useful to be able to derive $y \geq 0$ immediately without having to wait until the end of each path).
Yet, an SMT-solver based on DPLL(T) does not learn lemmas constructed from new predicates, while the approach in \parencite{Chu_Jaffar_EMSOFT2011} learns new predicates on-the-fly from Craig interpolants.
In our approach, we help the SMT-solver by preventively feeding ``candidate lemmas'', which, if used in a proof that there is no path longer than a certain bound, act as Craig interpolants, as explained in \autoref{sec:interpolants}.
Our approach therefore leverages both out-of-order predicate selection and interpolation, and, as a consequence, it seems to scale better. 
\smallskip

Two recent works --- \textcite{Biere_et_al_WCET2013} and its follow-up \textcite{DBLP:conf/rtns/KnoopKZ13} --- integrate the WCET path analysis into a counterexample guided abstraction refinement loop. 
As such, the IPET approach using ILP is refined by extracting a witness path for the maximal time, and testing its feasibility by SMT-solving; if the path is infeasible, an additional ILP constraint is generated, to exclude the spurious path. Because this ILP constraint relates all the conditionals corresponding to the spurious witness path, excluding infeasibile paths in this way exhibits an exponential behavior we strove to avoid. Moreover, our approach is more flexible with respect to (1) the class of properties which can be expressed, as it is not limited by the ILP semantics and (2) the ability to incorporate non-functional semantics (which is unclear whether \parencite{Biere_et_al_WCET2013} or \parencite{DBLP:conf/rtns/KnoopKZ13} can).

\textcite{metzner-CAV04} proposed an approach where the program control flow is encoded into a model along with either the concrete semantics of a simple model of instruction cache, or an abstraction thereof.
The WCET bound is obtained by binary search, with each test performed using the VIS model-checker\footnote{\url{http://vlsi.colorado.edu/~vis/}}.
\textcite{DBLP:conf/wcet/HuberS09} proposed a similar approach with the model-checker UPPAAL.\footnote{\url{http://www.uppaal.org/}}
In both cases, the functional semantics are however not encoded, save for loop bounds: branches are chosen nondeterministically, and thus the analysis may consider infeasible paths.
\textcite{dalsgaard-et-WCET10} encode into UPPAAL precise models of a pipeline, instruction cache and data cache, but again the program is modeled as ``data insensitive'', meaning that infeasible paths are not discarded except when exceeding a loop bound.


\textcite{Holsti_WCET2008} considers a loop (though the same approach can also be applied to loop-free code): the loop is sliced, keeping only instructions and variables that affect control flow, and a global ``timing'' counter $T$ is added; the input-output relation of the loop body is obtained as a formula in linear integer arithmetic (Presburger arithmetic); some form of \emph{acceleration} is used to establish a relation between $T$, some induction variables and some inputs to the program.
Applied to loop-free programs, this method should give exactly the same result as our approach. Its main weakness is that representations of Presburger sets are notoriously expensive, whereas SMT scales up (the examples given in the cited article seem very small, taking only a few lines and at most 1500 clock cycles for the entire loop execution); also, the restriction to Presburger arithmetic may exclude many programs, though one can model constructs outside of Presburger arithmetic by nondeterministic choices.
Its strong point is the ability to precisely deal with loops, including those where the iteration count affects which program fragments are active.

\section{Extensions and Future Work}
\label{sec:extensions}
The ``counter encoding'' is best suited for code portions that have a single entry and exit point (in which case they express the timing difference between entry and exit). In contrast, the ``sum encoding'' may be applied to arbitrary subsets of the code, which do not in fact need to be connected in the control-flow graph. One may thus use other heuristic criteria, such as usage of related variables.

A model based on worst-case execution times for every block, to be reassembled into a global worst-case execution time, may be too simplistic: indeed, the execution time of a block may depend on which blocks were executed beforehand, or, for finer modeling, on the value of pointer variables (for determining cache status).

A very general and tempting idea, as suggested earlier in MDD-based model-checking \parencite{metzner-CAV04}, in symbolic execution and bounded model checking by \parencite{DBLP:journals/rts/ChattopadhyayR13,Chu_Jaffar_EMSOFT2011}, in combined abstract interpretation and SAT-solving \parencite{DBLP:conf/rtas/BanerjeeCR13} is to integrate in the same analysis both the non-functional semantics (e.g. caches) and the functional semantics; in our case, we would replace both the micro-architectural analysis (or part of it) and the path analysis by a single pass of optimization modulo SMT.
Because merely encoding the functional semantics and a simplistic timing model already led to intractable formulas, we decided to postpone such micro-architectural modeling until we had solved scalability issues.
We intend to integrate such \emph{non-functional} aspects into the SMT problem in future work.

Detailed modeling of the cache, pipeline, etc. may be too expensive to compute beforehand and encode into SMT. One alternative is to iteratively refine the model with respect to the current ``worst-case trace'': to each basic block one attaches an upper bound on the worst-case execution time, and once a worst-case trace is obtained, a trace analysis is run over it to derive stronger constraints. These constraints can then be incorporated in the SMT encoding before searching for a new longest path.

We have discussed obtaining a tight upper bound on the worst-case operation time of the program from upper bounds on the execution times of the basic blocks. If using lower bounds on the worst-case execution times of the basic blocks, one may obtain a lower bound on the worst-case execution time of the program. Having both is useful to gauge the amount of over-approximation incurred. Also, by applying minimization instead of maximization, one gets bounds on best-case execution time, which is useful for some scheduling applications~\parencite{Wilhelm06}.

On a more practical angle, our analysis is to be connected to analyses both on the high level specification (e.g. providing invariants) and on the object code (micro-architectural timing analysis); this poses engineering difficulties, because typical compilation framework may not support sufficient tracing information.

Our requirement that the program should be loop-free, or at least contain loops with small constant bounds, can be relaxed through an approach similar to that of \textcite{Chu_Jaffar_EMSOFT2011}: the body of a loop can be summarized by its WCET, or more precisely by some summaries involving the cost variables and the scalar variables of the program. Then, this entire loop can be considered as a single block in an analysis of a larger program, with possibly overapproximations in the WCET, depending on how the summaries are produced.

\section{Conclusion}
\label{sec:conclusion}
We have shown that optimization using satisfiability modulo theory (SMT) is a workable approach for bounding the worst-case execution time of loop-free programs (or programs where loops can be unrolled). To our knowledge, this is the first time that such an approach was successfully applied.

Our approach computes an upper bound on the WCET, which may or may not be the actual WCET.
The sources of discrepancy are
\begin{enumerate*}[label={\arabic*)}]
\item the microarchitectural analysis (e.g. the cache analysis does not know whether an access is a hit or a miss),
\item the composition of WCET for basic blocks into WCET for the program, which may lose dependencies on execution history%
\footnote{This does not apply to some simple microcontroller architectures, without cache or pipeline states, e.g. Atmel AVR{\RegTM} and Freescale{\texttrademark} HCS12.},
\item the encoding of the program into SMT, which may be imprecise (e.g. unsupported constructs replaced by nondeterministic choices).
\end{enumerate*}

We showed that straightforward encodings of WCET problems into SMT yield problems intractable by all current production-grade SMT-solvers (``diamond formulas''), and how to work around this issue using a clever encoding.
We believe this approach can be generalized to other properties, and lead to fruitful interaction between modular abstraction and SMT-solving.

From a practical point of view, our approach integrates with any SMT solver without any modification, which makes it convenient for efficient and robust implementation. It could also integrate various simple analyses for introducing other relevant cuts.

While our redundant encoding brings staggering improvements in analysis time, allowing formerly intractable problems to be solved under one minute, the improvements in the WCET upper bound brought by the elimination of infeasible paths depend on the structure of the program being analyzed.
The improvement on the WCET bound of some industrial examples (18\%, 55\%\dots) is impressive, in a field where improvements are often of a few percents.
This means that, at least for certain classes of programs, it is necessary to take infeasible paths into account.
At present, certain industries avoid using formal verification for WCET because it has a reputation for giving overly pessimistic over-estimates; it seems likely that some of this over-estimation arises from infeasible paths.

Our approach to improving bounds on WCET blends well with other WCET analyses. It can be coupled with an existing micro-architectural analysis, or part of that analysis may be integrated into our approach. It can be combined with precise, yet less scalable analyzes \parencite{DBLP:conf/rtns/KnoopKZ13,Holsti_WCET2008} to summarize inner loops; but may itself be used as a way to summarize the WCET of portion of a larger program.

\printbibliography
\end{document}